\documentclass[conference, onecolumn]{IEEEtran}
\IEEEoverridecommandlockouts
\usepackage[letterpaper, left=1.4in, right=1.4in, bottom=1in, top=1in]{geometry}
\usepackage{cite}
\usepackage{amsmath,amssymb,amsfonts}
\usepackage{textcomp}
\usepackage{xcolor}

\usepackage{epsfig}
\usepackage{latexsym}
\usepackage{euscript}
\usepackage{multirow}
\usepackage{subfigure}
\usepackage{graphics,eepic,epic,psfrag}
\usepackage[ruled,vlined]{algorithm2e}

\usepackage{graphicx, float, caption, array, multirow}

\usepackage[utf8]{inputenc}
\usepackage[english]{babel}

\renewcommand{\thetable}{\arabic{table}}
\newcolumntype{P}[1]{>{\centering\arraybackslash}p{#1}}

\def\BibTeX{{\rm B\kern-.05em{\sc i\kern-.025em b}\kern-.08em
    T\kern-.1667em\lower.7ex\hbox{E}\kern-.125emX}}
\begin{document}

\title{Flow-Packet Hybrid Traffic Classification for Class-Aware Network Routing}

\author{
	\IEEEauthorblockN{Sayantan Chowdhury\IEEEauthorrefmark{1}, Ben Liang\IEEEauthorrefmark{1}, Ali Tizghadam\IEEEauthorrefmark{2}, Ilijc Albanese\IEEEauthorrefmark{2}}
	\IEEEauthorblockA{\IEEEauthorrefmark{1}University of Toronto, Canada \quad \IEEEauthorrefmark{2}TELUS, Canada}
}

\maketitle

\begin{abstract}
	Network traffic classification using machine learning techniques has been widely studied. Most existing schemes classify entire traffic flows, but there are major limitations to their practicality. At a network router, the packets need to be processed with minimum delay, so the classifier cannot wait until the end of the flow to make a decision. Furthermore, a complicated machine learning algorithm can be too computationally expensive to implement inside the router. In this paper, we introduce flow-packet hybrid traffic classification (FPHTC), where the router makes a decision per packet based on a routing policy that is designed through transferring the learned knowledge from a flow-based classifier residing outside the router. We analyze the generalization bound of FPHTC and show its advantage over regular packet-based traffic classification. We present experimental results using a real-world traffic dataset to illustrate the classification performance of FPHTC. We show that it is robust toward traffic pattern changes and can be deployed with limited computational resource.
\end{abstract}

\section{Introduction}
\label{sec:intro}

\begin{figure*}[t]
	\centering
	\includegraphics[width=14cm]{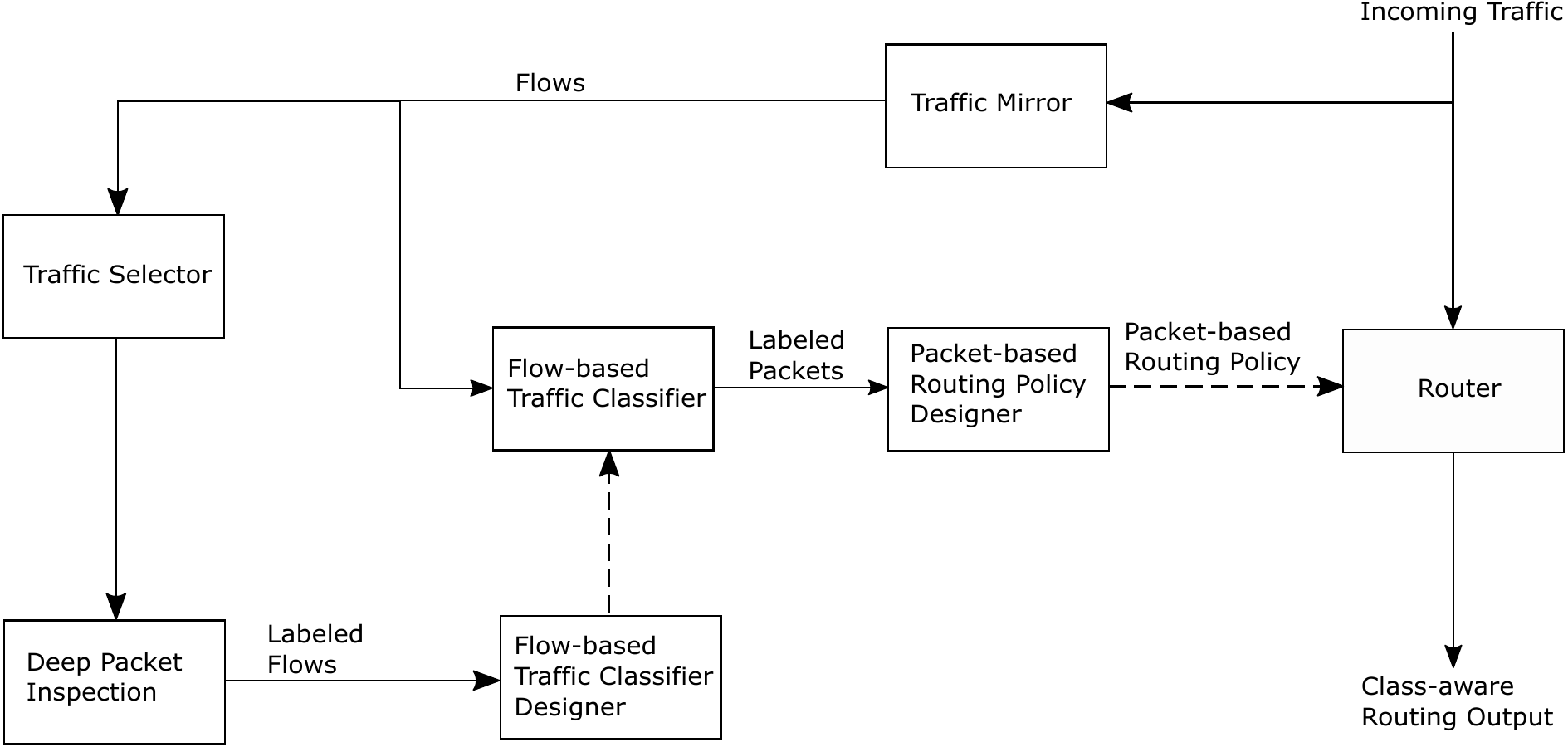}
	\caption{Diagram illustrating the FPHTC framework.}
	\label{fig:scheme}
\end{figure*}

Traffic classification is critical to the operation of computer networks in many aspects, such as network management, quality-of-service guarantee, and security concerns. As a large segment of network traffic is encrypted in recent years, traditional methods of classification, e.g., the protocol-based approach and content comparison, are no longer appropriate \cite{Finsterbusch14}. In contrast, machine learning algorithms can identify traffic flows with high accuracy using statistical features \cite{Moore05,Auld07,Duffield04,Wang16, Chowdhury19}.

For class-aware routing in a network, the routers need to conduct traffic classification before forwarding the traffic. However, common machine learning algorithms are often too computationally expensive for the routers. For example, even though deep neural networks is known to provide accurate classification outcomes, they are computationally intensive to train, while they need to be updated frequently to adapt to the changing traffic pattern over time. More importantly, the statistical features required for flow-based traffic classification techniques often are not available for real-time classification. In general, statistical features such as the \textit{variance of packet length} are extracted at the end of each flow, after the lengths of all packets are observed. Therefore, they are not useful for a router that must route the packets of a flow as they arrive, with as little delay as possible. Some authors have proposed early recognition of traffic classes by observing a subset of packets from each flow \cite{Bernaille07,Nguyen12, Xie12}. However, this still might cause significant delay as a router generally needs to process millions of packets within a fraction of a second.

A naive alternative is pure packet-based traffic classification, where each packet is observed and classified immediately, i.e., the classifier does not wait for a stream of packets from a flow. The router can look at the simple features embedded in the packet header and make a quick decision per packet. Packet-based traffic classification requires only fast lookup of the packet headers and thus is amenable to practical implementation in high-speed routers. However, a key drawback of this approach is that the classification performance can be poor due to the absence of detailed statistical features that are available in flow-based classification. 

This motivates us to combine the advantages of both flow-based and packet-based traffic classification. We propose a novel Flow-Packet Hybrid Traffic Classification (FPHTC) method, where a low-complexity routing policy with packet classification at the router is designed with the assistance of a flow-based classifier that resides outside the router. To the best of our knowledge, there exists no prior work that considers this hybrid form between flow-based and packet-based methods for network traffic classification in the router.

Our contributions can be summarized as follows:
\begin{itemize}
	\item We propose FPHTC, which generates a low-complexity routing policy to be applied to the incoming packets at a router. The routing policy enables class-aware routing using only simple features, e.g., those that can be directly read from the packet header. We generate the routing policy by exploiting the knowledge learned by a highly accurate flow-based classifier residing outside the router. The routing policy is constructed as a decision tree trained using the packets labeled by the flow-based classifier. In FPHTC, we can employ the flow-based classifier to label any number of packets, and thus, the resulting routing policy can be highly accurate.
	\item We show that FPHTC can be deployed in an online learning setting, where a new routing policy is updated at the router whenever the performance of the current routing policy falls below a certain threshold due to changes in the traffic pattern. This is achieved by adaptively re-training the flow-based classifier and then the routing policy, upon receiving the feedback that routing policy update is required.
	\item We provide theoretical justification for the performance advantage of FPHTC over regular packet-based traffic classification, in terms of the generalization bound. This further enables exploring the trade-off between the cost of labeling data for training the flow-based classifier and the generalization bound of the routing policy.
	\item We conduct extensive experiments using an aggregate dataset of 43590 encrypted traffic flows from \cite{Vpn16} and \cite{Tor17}. We train gradient boosted tree models XGBoost \cite{Xgb16} and LightGBM \cite{Lgbm17} as the flow-based classifier and compare the performance of FPHTC with regular packet-based traffic classification for different training dataset sizes. We observe substantial  performance gain under FPHTC.
\end{itemize}

The rest of this paper is structured as follows. The concept of FPHTC is presented in Section \ref{sec:proposed}, where we describe different components of FPHTC, routing policy design, and update procedure in detail. Section \ref{sec:policy} provides an analytical comparison between FPHTC and regular packet-based traffic classification in terms of the generalization bound. In Section \ref{sec:results}, we present our experimental setup and classification performance of FPHTC. Section \ref{sec:conclusion} concludes the paper.

\section{Flow-Packet Hybrid Traffic Classification}
\label{sec:proposed}

We propose FPHTC for a router that needs to conduct class-aware traffic processing. In this section, we provide a detailed description of our scheme. A diagram illustrating the overall framework of FPHTC is given in Fig.~\ref{fig:scheme}.

\subsection{Core Components of FPHTC}
\subsubsection{Router}
The router accepts an incoming stream of packets and processes them according to their service classes using the routing policy. The basic structure and function of such a routing policy are well-defined in prior works on packet classification \cite{Gupta99, Gupta01}. Throughout our work, we focus on how to generate routing
policy rules by training a machine learning model for packet-based traffic classification, where the chosen header fields of each packet are its features, i.e., the inputs into the learning model, and the packet is classified by the learning model to determine its CoS. For example, the chosen header fields may be the source IP address, destination IP address, source port number, and destination port number, among others, and the possible actions may be to route a packet as delay sensitive, delay moderate, or delay tolerant.

\subsubsection{Flow-based Traffic Classifier}
The flow-based traffic classifier resides outside the router, in some powerful equipment that can handle the heavy computation required by sophisticated machine learning techniques. It is a complex and highly accurate machine learning model that can classify a traffic flow in terms of CoS for all of its packets. It is trained using a number of bidirectional TCP flows with a set of flow-level statistical features extracted from the raw dataset.

Various methods are possible to generate the training dataset for the flow-based traffic classifier. In this work, since we are ultimately interested in online classification to handle changing traffic pattern over time, we propose to use a continuously updated recording of the past traffic. Specifically, we use a traffic mirror and a traffic selector, as shown in Fig.~\ref{fig:scheme}, to separate a selected small portion of the incoming traffic flows. The selected flows are then labeled using a Deep Packet Inspection (DPI) module according to their CoS. The true CoS labels obtained by DPI are used to train the flow-based classifier. We note that DPI cannot be used to replace the role of the flow-based classifier for all flows, due to its prohibitive cost and delay for common encrypted traffic. 

The role of the flow-based traffic classifier designer includes data preprocessing, hyperparameter selection, and finally, training the flow-based classifier. Once the flow-based classifier is trained, we use it to infer the CoS labels of all incoming flows captured by the traffic mirror. Then all packets belonging to a flow can be tagged by CoS label of the flow. We note that the CoS labels generated in this way, by a flow-based classifier, are too late to be used in the \textit{routing} of the labeled packets. However, what this achieves is to create a packet-level dataset for \textit{training} the packet-based routing policy as explained below.

\subsubsection{Packet-based Routing Policy Designer}

The packet-based routing policy designer takes labeled packets from the flow-based classifier as input, and it outputs a routing policy for the router. Specifically,  the routing policy designer trains a packet-based classifier using the labeled packets as the training dataset. 

In this work, we use the binary decision tree learning model for the packet-based classifier. In the decision tree, each path from the root to a node is a routing policy rule. Thus, to obtain routing policy rules that can be used in the router, the routing policy designer only needs to train a decision tree on the packet-level dataset. Furthermore, we note that the number of routing policy rules equals the number of leaf nodes in the decision tree. This provides an easy way to control the size of the routing policy, i.e., the routing policy designer can limit the maximum number of leaf nodes while training the decision tree.

\subsection{Construction of Routing Policy}

The construction of the routing policy in FPHTC involves transferring learned knowledge from the flow-based classifier to the routing policy designer. In the machine learning literature, knowledge distillation \cite{Hinton15, Vapnik16} is a technique where a simple student model is trained on the predictions supplied by a highly accurate and complex teacher model. In FPHTC, we train a decision tree at the routing policy designer using the predictions from the flow-based classifier as training targets. In essence, the routing policy designer tries to approximate the performance of the flow-based classifier. 

The flow-based classifier is trained with flow-level statistical features whereas the routing policy designer uses only some features that can be read directly from the packet header. Therefore, it is clear that the learned routing policy will perform worse than the flow-based classifier given the same traffic data for training. However, since there are unlabeled training data available, i.e., those that have not been labeled by DPI, we can label those data samples using our flow-based classifier to substantially enlarge the training dataset for the routing policy designer. Since the decision tree at the routing policy designer is trained on a much larger dataset than that of the flow-based classifier, the performance of the routing policy can be close to that of the flow-based classifier. More importantly, since the routing policy created by FPHTC utilizes information learned from a more powerful flow-based classifier, it can substantially outperform a regular packet-based classifier trained using only the small amount of labels generated by DPI.

\subsection{Routing Policy Update Procedure in Online Setting}

In a practical system, the data pattern of the incoming traffic changes over time, e.g., due to new applications appearing in the network, or changing user behavior. Therefore, we design FPHTC to dynamically update the routing policy over time.

\begin{figure}[t]
	\centering
	\includegraphics[width=9cm]{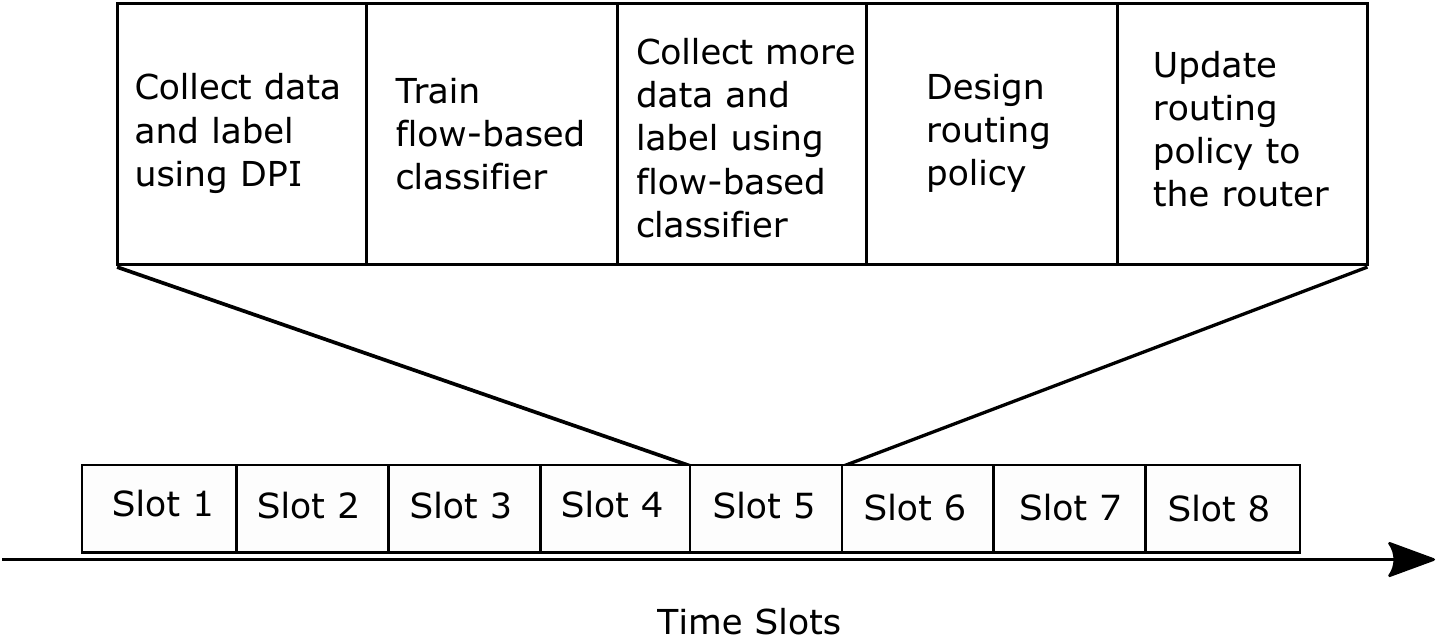}
	\caption{FPHTC in online setting.}
	\label{fig:online}
\end{figure}

In Fig.~\ref{fig:online}, we illustrate how the modules sequentially function over a continuous stream of traffic. At any given time slot, we collect and label a small portion of the incoming traffic flows using DPI to train the flow-based classifier. Meanwhile, we continue to collect flows to be used in the training of the routing policy. Once the flow-based classifier is trained, we use it to label those collected flows not labeled by DPI. Then, the routing policy designer trains a decision tree to generate the routing policy, which is then updated to the router. 

One important question is whether we should repeat these steps and update the routing policy at each time slot. If the traffic data pattern does not change too frequently, routing policy update at every time slot would be a waste of resources. To re-train the flow-based classifier, the labeling cost using DPI would also be expensive. A cost-effective solution is to update the routing policy only when the traffic pattern has altered significantly. This can be inferred by measuring the performance deterioration at the router. A feedback signal can be generated, for example, based on the increase in packet drop or congestion, to indicate that a routing policy update is necessary. We demonstrate the adaptiveness of FPHTC in the online setting in Section \ref{sec:results}.

\section{Comparison With Regular Packet-based Traffic Classification}
\label{sec:policy}

To highlight the benefit of combining flow-based and packet-based traffic classification in FPHTC, we compare it against regular packet-based traffic classification without the help from a flow-based classifier. In this section, we provide a theoretical justification of why FPHTC performs better than regular packet-based traffic classification.

In regular packet-based traffic classification, a decision tree at the routing policy designer is trained with the true labels of the packets. In contrast, in FPHTC only the flow-based classifier requires true labels.  Hence, for fair comparison, we maintain that the training dataset size of the regular packet-based traffic classifier is equal to the training dataset size of the flow-based classifier in FPHTC, when both are measured in terms of the number of flows.

Let $n$ be the number of flows in the training dataset for the routing policy designer in FPHTC. Recall that these flows are labeled by the flow-based classifier. The flow-based classifier is trained with a $\lambda$ fraction of these flows, which have been labeled by DPI. If the cost of labeling each flow using DPI is $c_\text{DPI}$, then the total cost is $n\lambda c_\text{DPI}$. Increasing $\lambda$ will lead to a more accurate flow-based classifier in FPHTC and ultimately a more accurate routing policy. However, this will also result in a greater cost of labeling flows. Thus, there exists a trade-off that should be carefully analyzed.

Suppose $\mathcal{H}$ is the hypothesis set of a classifier with some capacity measure $|\mathcal{H}|_C$. If $\hat{f}\in\mathcal{H}$ is the function learned by the classifier from $n$ training samples, and $f$ is the ground truth, i.e., the target function of interest, then the generalization bound can be expressed as follows \cite{Vapnik16}:
\begin{equation}
R(\hat{f})-R(f)\leq O\left(\frac{|\mathcal{H}|_C}{n^r}\right) + \epsilon,
\label{eq:g}
\end{equation}
where $R(\cdot)$ is the expected loss, the $O(\cdot)$ term is the estimation error, and $\epsilon$ is the approximation error. The rate of learning is given by $O(n^{-r})$. For difficult or \textit{non-separable} problems, $r=\frac{1}{2}$ and this represents a slow rate of learning. In contrast, for easy or \textit{separable} problems, where the trained classifier makes no training error, $r=1$ and this represents a fast rate of learning \cite{Vapnik16}.

In FPHTC, the flow-based classifier and the routing policy designer play the role of the teacher and the student respectively. Let $\mathcal{H}_\text{fl}$ be the hypothesis set for the flow-based classifier, with capacity measure $|\mathcal{H}_\text{fl}|_C$. Let $f_\text{fl}\in\mathcal{H}_\text{fl}$ be the function learned by the flow-based classifier and $f$ be the ground truth. Since only a fraction $\lambda$ of the flows are used for training, the generalization bound of the flow-based classifier is given by
\begin{equation}
R(f_\text{fl})-R(f)\leq O\left(\frac{|\mathcal{H}_\text{fl}|_C}{n\lambda}\right) + \epsilon_\text{fl},
\label{eq:t}
\end{equation}       
where $\epsilon_\text{fl}$ is the approximation error of the flow-based classifier. Here we use a common assumption that the rate of learning for the more sophisticated teacher is inversely proportional to the size of the training dataset, i.e., $O((n\lambda)^{-1})$. 

Similarly, let $\mathcal{H}_\text{rp}$ be the hypothesis set of the routing policy designer in FPHTC with capacity measure $|\mathcal{H}_\text{rp}|_C$, and let $f_\text{rp} \in \mathcal{H}_\text{rp}$ be the function determined by the routing policy designer.  We have   
\begin{equation}
R(f_\text{rp})-R(f_\text{fl})\leq O\left(\frac{|\mathcal{H}_\text{rp}|_C}{n^\alpha}\right) + \epsilon_\text{rp},
\label{eq:st}
\end{equation}
where $\epsilon_\text{rp}$ is the approximation error of the routing policy. As the student learns using the teacher's predictions, the decision boundary of the original classification problem has been translated to a smoother one. Thus, the student, aided by the teacher's predictions, learns at a faster rate than with the true labels, so that its rate of learning is represented by the parameter $0.5\leq\alpha\leq 1$. 

Combining (\ref{eq:t}) and (\ref{eq:st}), we get
\begin{equation}
\begin{split}
R(f_\text{rp})-R(f)&=R(f_\text{rp})-R(f_\text{fl})+R(f_\text{fl})-R(f)\\
&\leq O\left(\frac{|\mathcal{H}_\text{rp}|_C}{n^\alpha}\right) + \epsilon_\text{rp} + O\left(\frac{|\mathcal{H}_\text{fl}|_C}{n\lambda}\right) + \epsilon_\text{fl}\\
&\leq O\left(\frac{\lambda^\alpha|\mathcal{H}_\text{rp}|_C+|\mathcal{H}_\text{fl}|_C}{n^\alpha\lambda^\alpha}\right) + \epsilon_\text{rp}+ \epsilon_\text{fl}.
\end{split}
\label{eq:kd}
\end{equation}
This gives the generalization bound for FPHTC. We note that this bound improves as $\lambda$ increases. 

As a further step for system optimization, we can consider a weighted sum of the generalization bound and the DPI labeling cost as a function of $\lambda$:
\begin{equation}
C(\lambda) = K\cdot\frac{\lambda^\alpha|\mathcal{H}_\text{rp}|_C+|\mathcal{H}_\text{fl}|_C}{n^\alpha\lambda^\alpha} + \epsilon_\text{rp}+ \epsilon_\text{fl} + n\lambda c_\text{DPI},
\label{eq:cost}
\end{equation} 
where $K$ is a constant of proportionality. To minimize $C(\lambda)$, we differentiate (\ref{eq:cost}) wrt $\lambda$ and set the derivative equal to zero. Finally, we obtain the optimal $\lambda$ for FPHTC as
\begin{equation}
\lambda^* =\left(\frac{\alpha K|\mathcal{H}_\text{fl}|_C}{n^{1+\alpha} c_\text{DPI}}\right)^{\frac{1}{1+\alpha}}.
\label{eq:opt}
\end{equation}

In regular packet-based traffic classification, we have the same hypothesis set $\mathcal{H}_\text{rp}$, since it represents the capability of the same router. The function $f_\text{pk}\in\mathcal{H}_\text{rp}$ is chosen to approximate the ground truth $f$ without the help of the teacher. Again, we can bound the regular packet-based traffic classifier as follows:
\begin{equation}
R(f_\text{pk})-R(f)\leq O\left(\frac{|\mathcal{H}_\text{rp}|_C}{\sqrt{n\lambda}}\right) + \epsilon_\text{pk},
\label{eq:s}
\end{equation}
where $\epsilon_\text{pk}$ is the approximation error of the regular packet-based traffic classifier. As the student is trained using true labels in this case, the classification problem is difficult and the rate of learning is slower at $O((n\lambda)^{-1/2})$. Comparing (\ref{eq:s}) with (\ref{eq:kd}), we see that FPHTC outperforms regular packet-based traffic classification if the following inequality holds:
\begin{equation}
O\left(\frac{\lambda^\alpha|\mathcal{H}_\text{rp}|_C+|\mathcal{H}_\text{fl}|_C}{n^\alpha\lambda^\alpha}\right) + \epsilon_\text{rp}+ \epsilon_\text{fl}
\leq
O\left(\frac{|\mathcal{H}_\text{rp}|_C}{\sqrt{n\lambda}}\right) + \epsilon_\text{pk}.
\label{eq: ineq}
\end{equation} 

Let us now explain why it is reasonable for (\ref{eq: ineq}) to hold in our traffic classification problem. As the teacher is a highly complex flow-based classifier and the student is a simple decision tree trained at the routing policy designer, we have $|\mathcal{H}_\text{fl}|_C>>|\mathcal{H}_\text{rp}|_C$. Hence, FPHTC may be viewed as an instance of Hinton's knowledge distillation framework \cite{Hinton15}, except that in our case the feature spaces of the teacher and the student are different. Furthermore, the flow-based classifier is trained with many flow-level statistical features, whereas the routing policy is designed based on a much smaller number of packet-level features. Therefore, the approximation error of the routing policy is much larger than that of the flow-based classifier. Thus, in (\ref{eq: ineq}), $\epsilon_\text{fl}+\epsilon_\text{rp}<<\epsilon_\text{pk}$. Furthermore, as $\alpha\geq 0.5$, we have $n^\alpha\lambda^\alpha\geq \sqrt{n\lambda}$ when $\lambda\leq 1$. Therefore, even though $|\mathcal{H}_\text{fl}|_C>>|\mathcal{H}_\text{rp}|_C$, the large value of $|\mathcal{H}_\text{fl}|_C$ can be balanced by the parameters $\alpha$ and $\lambda$.

Note that $\alpha$ is an intrinsic parameter that we cannot control, but we can control $\lambda$. On the one hand, if $\lambda$ is close to 1, there is no benefit from using $\lambda$. On the other hand, if $\lambda$ is too small, both $n^\alpha\lambda^\alpha$ and $\sqrt{n\lambda}$ approaches zero and the difference between $\alpha$ and 0.5 is lost. However, a moderate $\lambda$ can satisfy (\ref{eq: ineq}) and allow FPHTC to outperform regular packet-based traffic classification.

\section{Experimental Evaluation}
\label{sec:results}

In this section, we first discuss our experimental setup. Then, we evaluate the performance of FPHTC and compare it against regular packet-based traffic classification to demonstrate the impacts of the training dataset size, and the online setting.

\subsection{Dataset and Learning Models}

We use the combined real-world traffic traces of ISCX VPN-nonVPN (2016) \cite{Vpn16, Vpn_url} and ISCX Tor-nonTor (2016) \cite{Tor17, Tor_url}. It contains pcap files for 43590 encrypted TCP bidirectional flows from 8 application types. We group these 8 application types into 3 CoS categories as shown in Table \ref{tab:cos}. For flow-based classification, we extract 268 features from each TCP flow. Our current set of features includes source and destination IP addresses in addition to the list of 266 features used in \cite{Chowdhury19}. 

We remove the flows with no payload from our dataset and use 90\% of the rest of the dataset as the full training dataset, plus 10\% for testing. In various experiments that require different training dataset sizes, we use a randomly selected subset of the full training dataset. All training datasets are balanced by applying the \textit{sklearn.utils.resample} function from scikit-learn v0.21.3 \cite{sklearn} so that the number of training samples in each CoS class is the same. All machine learning models are implemented in Python 3.8.3. We use balanced test accuracy as the main performance metric, which is the  average of the proportion of correctly classified samples in each class. This performance metric is not affected by the imbalance in the dataset.

The routing policy designer takes labeled packets as input with 4 features: source IP address, destination IP address, source port number, and destination port number. The 32-bit IPv4 addresses are converted to decimal numbers to be used as feature values. Given a set of labeled flows, we obtain the corresponding set of unique packets having these 4 features. Note that these 4 features guarantee that all packets from a particular TCP flow are mapped to the same routing output.

We note that the traffic dataset is structured. For this type of data, it is known that the gradient boosted tree ensemble is appropriate as a learning model \cite{Kdd15}. We use the state-of-the-art gradient boosted tree ensembles XGBoost \cite{Xgb16} and LightGBM \cite{Lgbm17} as the flow-based classifier. In our experiments, the XGBoost model is trained with 100 trees in the ensemble. The learning rate is 0.3 and the maximum depth of each tree is limited to 6. For LightGBM, we use the gradient-based one-side sampling (GOSS) boosting method. The number of trees in the ensemble is 100 with the number of leaves and maximum depth restricted to 31 and unlimited, respectively. The learning rate remains 0.3.  

The routing policy designer trains a single CART with the predictions from the flow-based classifier as training targets. We use a decision tree classifier with balanced class weights and entropy as the criterion for choosing the best split. The parameters such as maximum depth and maximum leaf nodes, which determine the structure of the tree, are kept unlimited.

\begin{table}[t]
	\centering
	\begin{tabular}{|P{0.4\textwidth}|P{0.4\textwidth}|}
		\hline
		CoS label       & Application type    \\ \hline
		Delay Sensitive & CHAT, VOIP          \\ \hline
		Delay Moderate  & AUDIO, VIDEO        \\ \hline
		Delay Tolerant  & FTP, MAIL, P2P, WEB \\ \hline
	\end{tabular}
	\caption{Application types and CoS labels.}
	\label{tab:cos}
\end{table}

\subsection{Impact of Size of Training Datasets}

\begin{figure}[t]
	\centering
	\includegraphics[width=10cm]{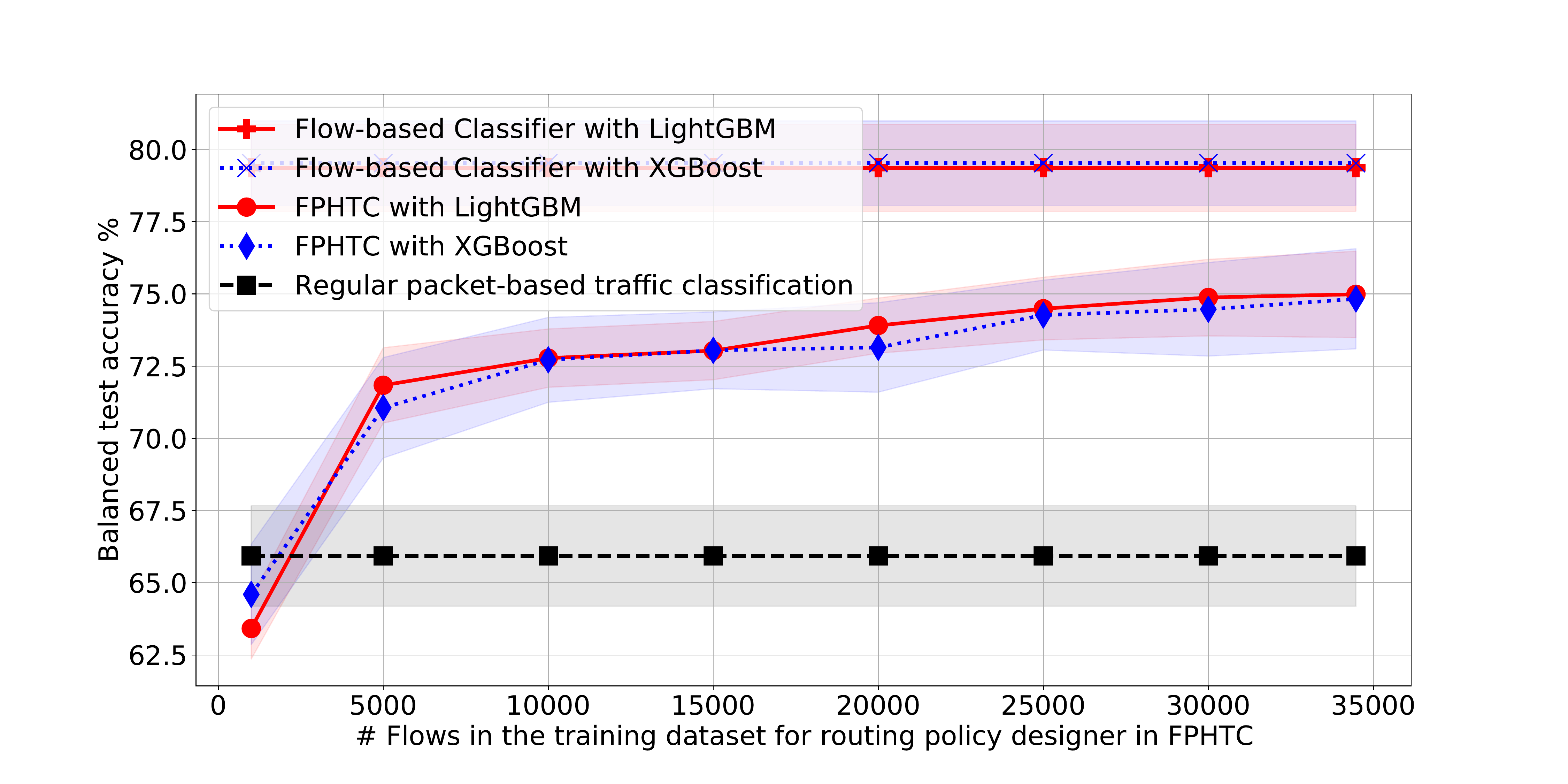}
	\caption{Balanced test accuracy vs.~$n$, size of training dataset for the routing policy designer in FPHTC, with 90\% confidence interval.}
	\label{fig:hybrid1}
\end{figure}

\begin{table}[t]
	\centering
	\begin{tabular}{|P{0.15\textwidth}|P{0.15\textwidth}|P{0.15\textwidth}|P{0.15\textwidth}|}
		\hline
		\# Flows in the training dataset & Flow-based classifier& FPHTC & Regular packet-based \\ \hline
		1000                                                         & 79.59\% & 74.99\%  & 65.93\%             \\ \hline
		5000                                                         & 89.59\%  & 84.85\%  & 78.97\%             \\ \hline
		10000                                                       & 92.27\% & 87.13\%  & 83.85\%             \\
		\hline
	\end{tabular}
	\caption{Balanced test accuracy vs.~\# flows in the training dataset for the flow-based classifier and the regular packet-based classifier.}
	\label{tab:flow-based}
\end{table}      

The performance of FPHTC is shown in Fig.~\ref{fig:hybrid1} with 1000 flows in the training dataset of the flow-based classifier. We present the balanced test accuracy of FPHTC with 90\% confidence interval. We observe that it increases as the training dataset size for the routing policy designer is increased. Recall that, since this training dataset is generated by the flow-based classifier, there is no theoretical limit to its size. In contrast, the training dataset size for the regular packet-based traffic classifier remains the same as that of the flow-based classifier. Thus, we observe that FPHTC outperforms regular packet-based traffic classification when the training dataset for the routing policy designer becomes large enough. The performance gain of FPHTC is larger when the flow-based classifier and the regular packet-based traffic classifier are trained using a small training dataset. For example, when the flow-based classifier with LightGBM and the routing policy designer are trained using 1000 flows and the maximum number of available flows, i.e., full training dataset containing 34473 flows, respectively, FPHTC is about 9\% more accurate than regular packet-based traffic classification. Thus, FPHTC significantly improves accuracy in the low data regime.

We note that the performance of FPHTC is strictly ascending in Fig.~\ref{fig:hybrid1}. If we had more than 34473 unique flows in our available dataset for training, we could have further increased the training dataset size for the routing policy designer and observed a larger gain of FPHTC over regular packet-based traffic classification. Also, we see that using LightGBM as the flow-based classifier performs slightly better than using XGBoost. Therefore, we will present the results using LightGBM as the flow-based classifier for the rest of the experiments.

In Table \ref{tab:flow-based}, for various training dataset sizes, we present the balanced test accuracy of the flow-based classifier with LightGBM, regular packet-based traffic classifier, and FPHTC, where $n=34473$. The experiment has been repeated over 10 randomly chosen training and test sets and then the average accuracy is listed. We note that there is a significant gap between the accuracy of the flow-based classifier and the regular packet-based traffic classifier. The gap is especially large when we have a small training dataset. This gap is due to the many more features observed by the flow-based classifier than the regular packet-based traffic classifier. This confirms our rationale for this work, that there is room for improvement for the routing policy if it can utilize the knowledge learned by the flow-based classifier. Furthermore, we observe that FPHTC can substantially reduce that gap, especially when the amount of training data is small.

\subsection{Classification Performance of FPHTC in Online Setting}

Finally, we implement the routing policy update procedure of FPHTC in an online setting. In this experiment, we simply use the test accuracy of the routing policy as a feedback signal. If the accuracy is dropped below some accuracy threshold at the end of a time slot, re-training of the flow-based classifier and the routing policy update begin from the next time slot. Training stops when the accuracy crosses back above the accuracy threshold and the consecutive improvement in accuracy is less than some saturation threshold.

\begin{figure}[t]
	\centering
	\includegraphics[width=10cm]{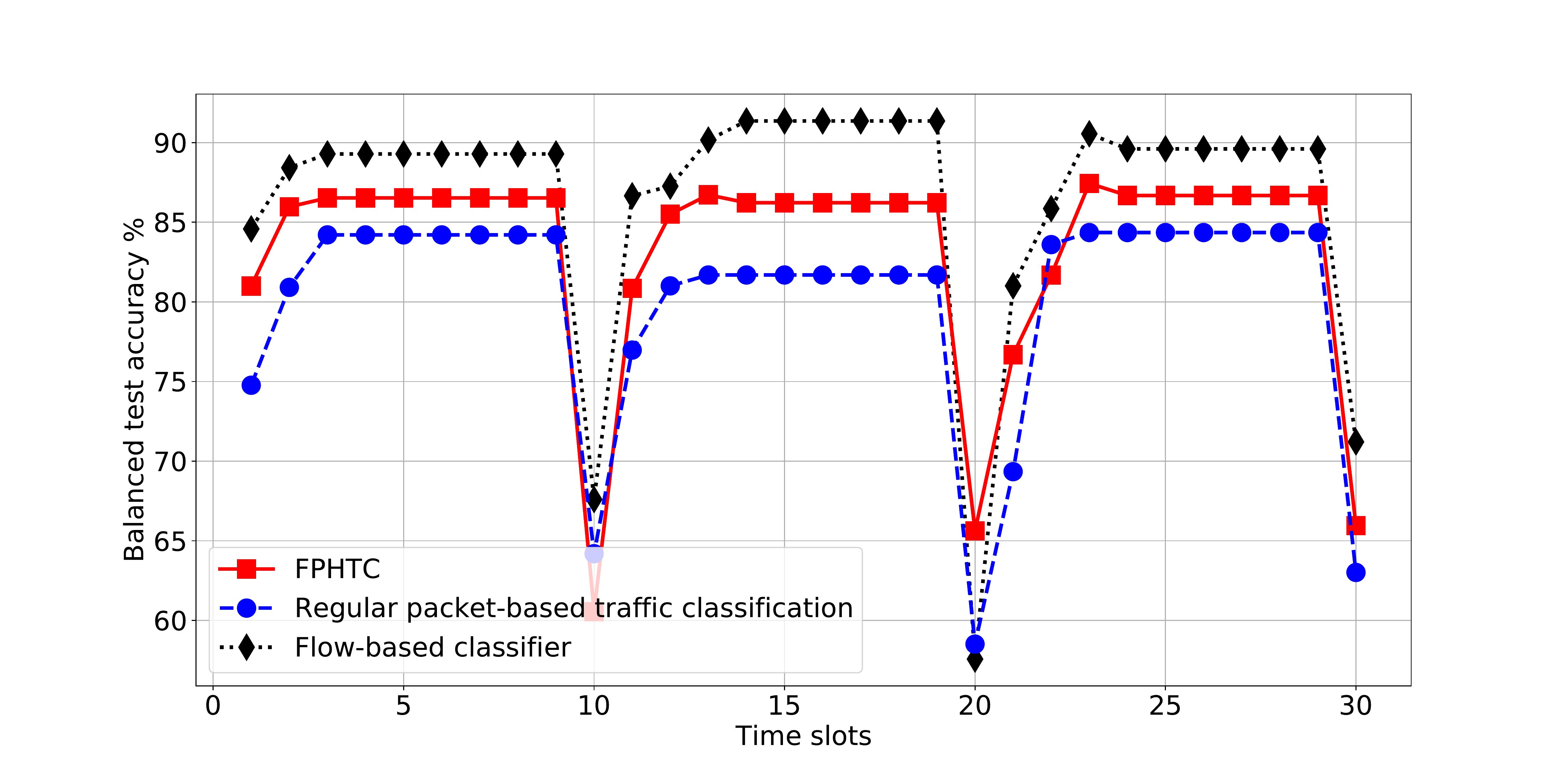}
	\caption{Performance of FPHTC in online setting with traffic pattern changing after every 10 time slots.}
	\label{fig:online_FPHTC}
\end{figure}

\begin{table}[t]
	\centering
	\begin{tabular}{|P{0.27\textwidth}|P{0.35\textwidth}|}
		\hline
		& Application types in the dataset    \\ \hline
		At time slot 0 & AUDIO, FTP, VIDEO, VOIP, WEB          \\ \hline
		At time slot 10  & FTP, MAIL, P2P, VIDEO, VOIP      \\ \hline
		At time slot 20  & AUDIO, CHAT, FTP, MAIL, WEB\\ \hline
	\end{tabular}
	\caption{Change of traffic pattern over time.}
	\label{tab:pattern}
\end{table}

To simulate a never-ending stream of traffic, we keep shuffling our training dataset randomly. We run our experiment over 30 time slots, and we change the data pattern after every 10 time slots. To simulate traffic pattern change, we always use only a subset of 5 application types as the incoming traffic. After 10 time slots, we pick another subset of 5 application types and generate the incoming traffic. Thus, the test traffic pattern is changed at time slot 0, at time slot 10, and at time slot 20. The changing of applications over time is shown in Table \ref{tab:pattern}. 

At each time slot where re-training is needed, the first 1000 flows are selected and labeled by DPI to be used for training the flow-based classifier. Then, 10000 flows are labeled by the trained flow-based classifier and used for routing policy design. We use an accuracy threshold of 80\% and a saturation threshold of 1\% in our experiment. Fig.~\ref{fig:online_FPHTC} illustrates how the balanced test accuracy drops after every 10 time slots due to the traffic pattern change. Then re-training begins, the routing policy is updated, and we observe gradual increase of the test accuracy. Once the test accuracy is saturated, the routing policy update stops, and we observe a flat line in the test accuracy. Again, we observe that FPHTC substantially outperforms regular packet-based traffic classification in the online setting.

\section{Conclusion}
\label{sec:conclusion}
In this paper, we propose Flow-Packet Hybrid Traffic Classification (FPHTC), which enables low-complexity but highly accurate packet-based classification at a network router. In FPHTC, a sophisticated flow-based classifier
residing outside the router uses high-dimensional flow-level features to achieve high classification accuracy, while a routing policy designer generates a simple routing policy for the router based on a small number of packet-level features, utilizing the knowledge of the flow-based classifier. We discuss routing policy updates in an online setting, which keeps FPHTC robust towards traffic pattern change over time. Our experimental results confirms that the learned knowledge from the flow-based classifier can substantially improve the performance of the routing policy.

\bibliography{references}

\bibliographystyle{IEEEtran}

\end{document}